# Accurate Mode-Coupling Characterization of Low-Crosstalk Ring-Core Fibers using Integral Calculation based Swept-Wavelength Interferometry Measurement

Junwei Zhang, Jiangbo Zhu, Junyi Liu, Shuqi Mo, Jingxing Zhang, Zhenrui Lin, Lei Shen, Lei Zhang, Jie Luo, Jie Liu, and Siyuan Yu

*Abstract*—In this paper, to accurately characterize the low inter-mode coupling of the weakly-coupled few mode fibers (FMFs), we propose a modified inter-mode coupling characterization method based on swept-wavelength interferometry measurement, in which an integral calculation approach is used to eliminate significant sources of error that may lead to underestimation of the power coupling coefficient. Using the proposed characterization method, a low-crosstalk ring-core fiber (RCF) with low mode dependent loss (MDL) and with single span length up to 100 km is experimentally measured to have low power coupling coefficients between high-order orbital angular momentum (OAM) mode groups of below -30 dB/km over C band. The measured low coupling coefficients based on the proposed method are verified by the direct system power measurements, proving the feasibility and reliability of the proposed inter-mode coupling characterization method.

## I. INTRODUCTION

Multi-mode optical systems based on optical fibers that support modes or mode groups (MGs) with desirable characteristics have become an important research topic in both classical and quantum photonic information systems. Mode-division multiplexing (MDM), which utilizes multiple optical modes in one guiding fiber core as independent data communication channels to provide high communication capacity density, has been widely considered as one of the promising solutions to overcome the nonlinear Shannon limit of conventional single-mode fiber (SMF) based communication systems [1], [2]. The inevitable inter-mode or inter-MG coupling in MDM fibers is a key factor in limiting the performance and scalability of such systems [3]. For example, attempts to up-scale MDM systems by introducing more mode channels are often limited by the fact that the inter-mode coupling (i.e., crosstalk) necessitates multiple-input multiple-output (MIMO) digital signal processing (DSP), and the DSP complexity increases dramatically as more modes and larger differential mode group delays (DMGDs) are involved [4], incurring unviable costs and power consumption. Thus design and fabrication of few mode fibers (FMFs) with low inter-mode or inter-MG coupling are crucial to realize low-complexity data transmission over long MDM fibers. Meanwhile, experimental characterization of low levels of inter-mode or inter-MG power coupling coefficients with high accuracy is also desirable, since it can exactly reflect the feasibility of the fiber design method and may reveal more details of the physical mechanisms underlining inter-mode coupling, therefore pointing to paths of further suppression of inter-mode coupling.

Several characterization methods of fiber inter-mode or inter-MG coupling have been reported, such as those based on swept-wavelength interferometry (SWI) [5]-[7], vector network analyzer (VNA) [8]-[10], and direct system power measurement [11]-[15]. However, the measured results out of such schemes are often highly dependent on the device performance used in the measurement setup and there exists non-negligible gaps among results from different methods for MDM fibers with similar refractive index profile (RIP) or even one specific fiber. For example, the power coupling coefficient between MGs $LP_{11}$ and $LP_{21}$ in [10] can be deduced to be < -50 dB/km according to the measured time-domain impulse response matrix using VNA setup, which is much lower than that of > -14 dB/360m using the direct system power measurement. In addition, the characterized low inter-mode power coupling coefficients/ crosstalk of the fibers using the SWI or VNA setup in [5]-[10] have not been verified by direct

Manuscript received March X, 2021. This work is supported in part by National Key R&D Program of China (2018YFB1801800), NSFC-Guangdong joint program (U2001601), National Natural Science Foundation of China (61875233), The Key R&D Program of Guangdong Province (2018B030329001), and Local Innovative and Research Teams Project of Guangdong Pearl River Talents Program (2017BT01X121). (Corresponding author: *Jie Liu*)

J. Zhang, J. Liu, S. Mo, J. Zhang, Z. Lin, J. Liu and S. Yu are with the State Key Laboratory of Optoelectronic Materials and Technologies, School of Electronics and Information Technology, Sun Yat-Sen University, Guangzhou 510006, China (e-mail: liujie47@mail.sysu.edu.cn).
J. Zhu is with the Department of Mathematics, Physics and Electrical Engineering, Northumbria University, Newcastle upon Tyne NE1 8ST, UK.
L. Shen, L. Zhang and J. Luo are with the Yangtze Optical Fiber and Cable Joint Stock Limited Company, State key Laboratory of Optical Fiber and Cable Manufacture technology No.9 Guanggu Avenue, Wuhan, Hubei, China.

system power measurements or data transmission demonstrations. Therefore, a more reliable characterization method needs to be developed to provide more accurate power coupling coefficient values.

In this paper, to accurately characterize the inter-mode coupling of the weakly-coupled FMFs, a modified inter-mode coupling characterization method based SWI measurement is proposed, in which an integral calculation approach is used to extract the power coupling coefficient with high accuracy. A ring-core fiber (RCF) with low-crosstalk and low mode dependent loss (MDL) [14], [15] is utilized to evaluate the accuracy of proposed inter-mode coupling characterization method. The measured results of the proposed characterization method are verified by direct optical power measurement over 100-km long fiber span, eliminating significant sources of error that may lead to underestimation of the power coupling coefficient. Using this improved inter-mode coupling characterization method, low average power coupling coefficients of <-31 dB/km between adjacent high-order orbital angular momentum (OAM) MGs (topological charge $|l|$ = 1 & 2, 2 & 3) within the entire C band are achieved after transmission of low-crosstalk RCFs with single-span length up to 100 km.

The remainder of the paper is organized as follows. In section II, we describe the operating principle of the modified inter-mode coupling characterization method and the effect on the measurement accuracy in the presence of MDL. In section III, we discuss the experimental setup and results. Finally, Section VI summarizes and concludes this paper.

## II. PRINCIPLE OF THE MODIFIED INTER-MODE COUPLING CHARACTERIZATION METHOD

### A. Principle of the modified inter-mode coupling characterization method

The power coupling coefficient $h$ between a pair of adjacent high-order modes (mode 1 and mode 2) can be evaluated based on measured impulse responses after fiber transmission, when exciting either mode 1 or 2. As can be seen from the measured impulse responses in the weak mode coupling regime in Figs. 1(a) and 1(b), the power coupling coefficient $h$ could be obtained by fitting the theoretical curves of Eqs. (2)-(7) (see the appendix for more details) to the measured impulse responses from the two coupling modes/ MGs [8,16]. However, there are two significant sources of errors using such theoretical-curve fitting approach:

1) The received discrete peaks corresponding to the excited modes/ MGs cease to be delta function due to pulse broadening attributed to modal dispersion, phase noise and bandwidth limitation of the measurement system, as the measured impulse response shown in Fig. 1(b).

2) Measured power fluctuations within the range of DMGD resulting from distributed mode coupling in the fiber make the fitting highly unreliable, as shown in Fig. 1(a).

We therefore propose a modified power-coupling-coefficient estimation approach with much more reliability only based on the experimentally measured impulse responses, which can be called an integral calculation approach, whose detailed steps are:

1) Step 1: temporal integrals of the excited mode peak and the DMGD plateau, as the red areas depicted in Figs. 1(d) and 1(c), are respectively calculated based on the measured impulse responses. These correspond to the power in the excited mode and of the accumulated light coupling over the entire measured fiber length. Here note that the crosstalk from mode 1 to mode 2 caused by the mode multiplexer (Mux) and demultiplexer (DeMux) can be easily distinguished from that resulting from the fiber-mode weak coupling, due to their different temporal positions as illustrated in Fig. 1(c).

2) Step 2: ratio of integrals of the DMGD plateau and the excited mode peak is calculated, which represents the accumulated crosstalk ($XT$) between mode 2 and mode 1 of the whole measured fiber.

3) Step 3: average power coupling coefficient $h$ can then be calculated using the equation below [16]:

$$h = \frac{\text{arctanh}(XT)}{z} \quad (1)$$

where $z$ is the fiber length.

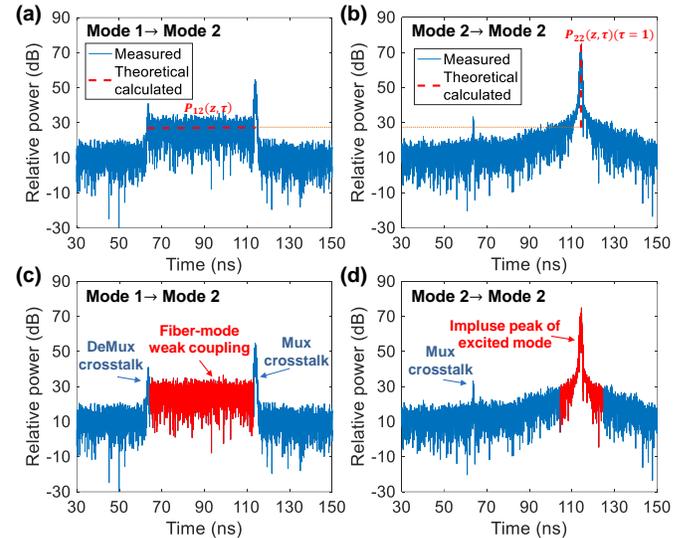

Fig. 1 The measured impulse responses in the weak mode coupling case for (a), (c) mode 1 excitation & mode 2 detection and (b), (d) mode 2 excitation & mode 2 detection. The red dotted lines and red areas represent the theoretical-fitting curves and the total power of strong launched mode/ mode coupling, respectively.

In this approach, the temporal integrals of the excited mode peak and plateau minimise errors in the theoretical curving fitting method where the amplitude of impulse peak and the plateau are used for $XT$ evaluation, as the peak amplitude is susceptible to variations caused by impulse broadening in the measurement system and the plateau is susceptible to noise. The calculation of power coupling coefficient $h$ based on Eq. (1) can also minimise errors caused by power fluctuations in the DMGD plateau.

The evaluation results obtained by this approach can also be double checked by direct crosstalk power measurement in an MDM system with cascaded mode Mux, MDM fiber and mode DeMux. For instance, the total system crosstalk introduced by both fiber-mode weak coupling and mode (De) Mux can also be evaluated by calculating temporal integrals of the whole time regions shown in Figs. 1(c) and 1(d) in Step 1 and then obtaining the XT in Step 2. To avoid power fluctuations due to multimode interference, broadband light source or modulated signal light should be adopted in the direct power measurement system.

The power coupling coefficients will be evaluated by the theoretical-curve fitting and proposed integral calculation approaches based on the measured impulse responses using the SWI [5]-[7] and VNA [8]-[10] setups, whose measurement accuracy will be verified by the direct system power measurement.

### B. Effect on the measurement accuracy in the presence of MDL

In previous section, the MDL is not considered since the proposed inter-mode coupling characterization method is suitable for short fiber span, in which the MDL is negligible and crosstalk caused by the mode Mux and DeMux can also be eliminated for the calculation of power coupling coefficient.

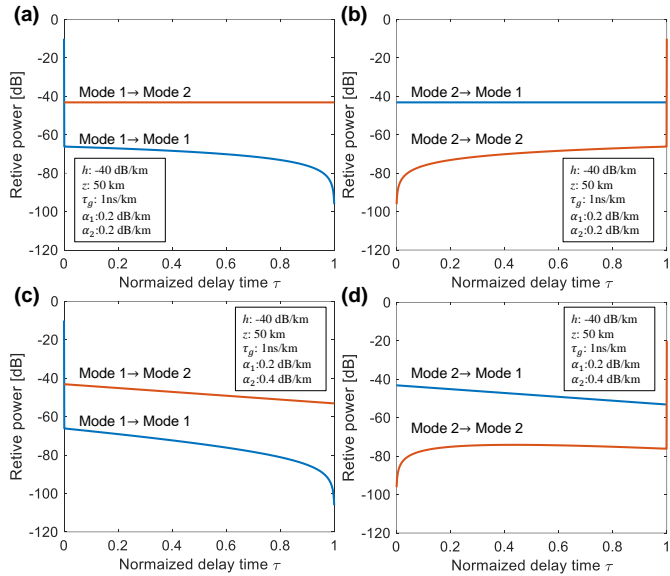

Fig. 2 Calculated impulse responses in the weak mode coupling case for (a) mode 1 excitation and (b) mode 2 excitation without MDL; calculated impulse responses in the weak mode coupling case for (c) mode 1 excitation and (d) mode 2 excitation with MDL.

The theoretical impulse responses in weak mode coupling case in the presence of MDL and the corresponding effect on the measurement accuracy of proposed method are analyzed. Figs. 2(a), 2(b) and 2(c), 2(d) show the calculated impulse responses without MDL and with MDL, respectively, based on Eqs. (2)-(7) (see the appendix for more details) when mode 1 or 2 is individually excited by impulse. Here we set $h$ = -40 dB/km, $z$ = 50 km, $\tau_g$ = 1 ns/km as in [8]. Besides, $\alpha_1 = \alpha_2$ = 0.2 dB/km and $\alpha_1$ = 0.2 dB/km, $\alpha_2$ = 0.4 dB/km are set for the calculations without MDL and with MDL, respectively. Compared with no MDL case, the two impulse responses corresponding to two excited modes with MDL have 10 dB power difference, due to the introduction of MDL after 50-km fiber transmission. Moreover, the non-excited mode power no longer uniformly distributes within the range of DMGD. Compared with mode 1, the mode 2 with a higher attenuation value suffers from a larger crosstalk value based on the output power difference using both integral calculation and direct system power approaches.

In order to avoid the effect of MDL on the measurement accuracy, the power coupling coefficient between modes or MGs can be calculated by summing or averaging the integral values of all excited mode peaks (corresponding to the blue line of Fig. 2(c) and the red line of Fig. 2(d)) and the integral values of all DMGD plateaus (corresponding to the red line of Fig. 2(c) and the blue line of Fig. 2(d)) belonging to the two measured modes or MGs in Step 1. Compared with direct system power measurement, the proposed integral calculation method can accurately evaluate the inter-mode coupling with distinguishable crosstalk from Mux/ DeMux.

## III. EXPERIMENTAL SETUP AND RESULTS

### A. Impulse Response Measurement Using VNA Setup

The cross-section and the refractive index profile (RIP) of the RCF used in the experiment setup (see Fig. 4) are depicted in Figs. 3(a) and (b), respectively. RIP modulation is implemented by placing an index notch within the ring-core area, which decreases the micro-perturbation induced inter-MG coupling [14], [15]. The characteristics of the used RCF including the effective refractive index difference ($\Delta n_{eff}$), propagation loss and DMGD are shown in Table I [14]. All MGs show a similar attenuation of around 0.21 dB/km with a low MDL of ≤ 0.003 dB/km.

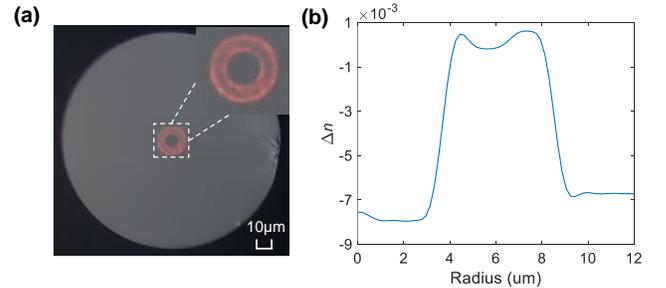

Fig. 3 (a) Cross-sectional diagram and (b) RIP of the RCF used in the experimental setup.

TABLE I. CHARACTERISTICS OF THE USED RCF.

|  |  | $|l|$ = 0 | $|l|$ = 1 | $|l|$ = 2 | $|l|$ = 3 |
|---|---|---|---|---|---|
| $\Delta n_{eff}$ | $\times 10^{-3}$ | 0.7 | 1.8 | | 2.5 |
| Propagation Loss | dB/km | 0.209 | 0.211 | 0.208 | 0.210 |
| DMGD to MG $|l|$ = 0 | ns/km | 0 | 3.2 | 8.5 | 13.1 |

The impulse responses involving adjacent OAM MGs are experimentally measured using the VNA setup [8] as shown in Fig. 4. The radio frequency (RF) signals with a frequency sweeping from 20 MHz to 20/ 5 GHz from the VNA (operated in impulse response mode) drive a Mach-Zehnder modulator (MZM) to produce the optical signal for 1-km/ 10-km measurement. The optical carrier is generated by a tunable laser at 1550 nm. After amplification by an erbium doped fiber amplifier (EDFA), the optical signal is converted to OAM beam by a spatial light modulator (SLM). The OAM beam is converted to circular polarization state by a quarter-wave plate (QWP) before being collimated into the 1-km or 10-km RCF. After fiber transmission, all output modes are simultaneously coupled into a multi-mode fiber (MMF) pigtailed photo detector (PD). Finally, the detected RF signals are fed back to the VNA for impulse response measurement. Only the impulse responses over the 1-km and 10-km RCF span are measurable in this approach. Due to the low received optical power and bandwidth limitation, the same measurement cannot be attained in the 100-km RCF span.

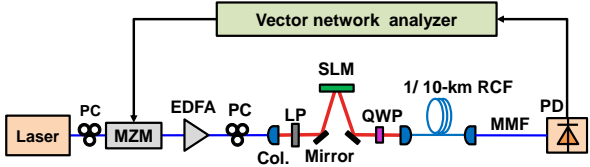

Fig. 4 VNA-based impulse response measurement setup for determining mode coupling in a fiber. PC: polarization controller; MZM: Mach-Zehnder modulator; EDFA: erbium doped fiber amplifier; Col.: collimator; LP: linear polarizer; SLM: spatial light modulator; QWP: quarter-wave plate; PD: photo detector.

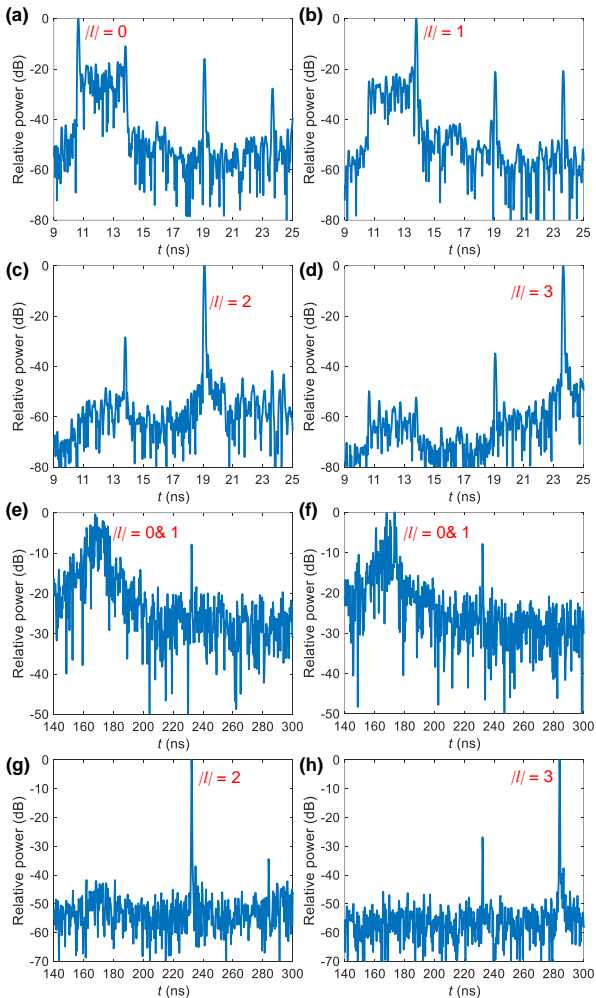

Fig. 5 The VNA-measured impulse responses at 1550 nm, with launched MGs of (a) $|l| = 0$, (b) $|l| = 1$, (c) $|l| = 2$ and (d) $|l| = 3$ over 1-km RCF and with launched MGs of (e) $|l| = 0$, (f) $|l| = 1$, (g) $|l| = 2$ and (h) $|l| = 3$ over 10-km RCF.

The measured impulse responses for all supported OAM MGs at 1550 nm using VNA setup over 1-km and 10-km RCFs are shown in Figs. 5(a)–(d) and 5(e)–(h), respectively. Relatively strong distributed mode coupling between OAM MGs $|l| = 0$ and $|l| = 1$, due to their small $\Delta n_{eff}$, is observed as one very wide Gaussian-shape peak after 10-km RCF transmission. However, the deduced power coupling coefficient between MGs $|l| = 0$ and $|l| = 1$ for 1-km VNA-measured impulse responses using the integral calculation and theoretical-curve fitting approaches are < -10 dB/km and < -20 dB/km, respectively, which conflict with the strong coupling results of 10-km measurement system.

Meanwhile, the DMGD plateaus resulting from distributed weak coupling between high-order OAM MGs with topological charge of $|l| > 1$ are extremely low for both 1-km and 10-km systems. Compared with 1-km case, the DMGD plateaus are almost hidden below the noise floor of the measurement system after 10-km transmission due to reduction of signal-to-noise ratio with direct detection over a long fiber. These results indicates that the VNA method somehow suppresses the manifestation of weak inter-MG coupling induced by random perturbations and produces a significant under-estimation of the power coupling coefficient similar as the measured impulse responses shown in [9], [10]. The reason of this phenomenon is analyzed as follows.

Although the VNA is operated in the impulse response mode, the actual measurement is a radio-frequency (RF) frequency-domain response measurement, with the obtained frequency response converted to impulse response by means of an internal inverse fast Fourier transform (IFFT) algorithm. This method is suitable for measuring discrete coupling such as those at the Mux/ DeMux and to a good extent for strong coupling as the response approaches a single impulse. In the distributed weakly coupled regime, only coupling from the strong launched mode to the weak destination mode (the mode being measured) needs to be considered. As the VNA's dwelling time at each RF signal frequency is much longer than the total group delay of the fiber (~ 50 us/ 10 km), the response obtained at each RF frequency is the sum of all RF signals coupled from the launched mode to the mode being measured along the entire length of the fiber. Due to the DMGD between the two modes, the RF signals coupled over different fiber positions will have different phase. Therefore, the summation of the RF signals amounts to an integral over the total RF phase shift accrued over the range of DMGD. The $\Delta n_{eff}$ value of $1.8 \times 10^{-3}$ is equivalent to an inter-MG DMGD of ~ 50 ns over the 10-km length, which is longer than the RF signal period $T_{RF}$ for frequencies of > 20 MHz, giving rise to a total RF phase shift distributed over more than $[0, 2\pi]$. Assuming a uniform coupling over the entire length and ignoring the small fiber attenuation (~ 0.21 dB/km), the RF response associated with the destination mode as measured at the end of the fiber would have been averaged to a small value close to zero. This of course does not mean that mode coupling has not happened, but the mere fact that the RF components coupled from the strong launched mode have been averaged out, leaving only the RF components of the strong launched mode that can be detected by the VNA.

## B. Impulse Response Measurement Using SWI Setup

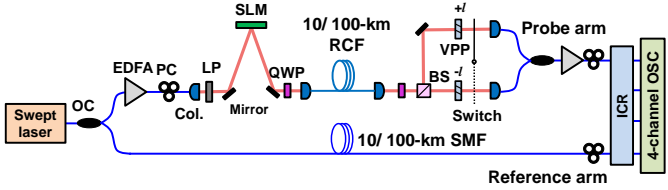

Fig. 6 SWI-based impulse response measurement setup for determining mode coupling in a fiber. EDFA: erbium doped fiber amplifier; PC: polarization controller; Col.: collimator; LP: linear polarizer; SLM: spatial light modulator; QWP: quarter-wave plate; BS: beam splitter; VPP: vortex phase plate; ICR: integrated coherent receiver; OSC: oscilloscope.

Time-domain impulse response obtained using a SWI setup [5]-[7] can also be used to evaluate the inter-MG power coupling coefficient. The RCFs under test were continuously drawn from the same preform before cut into 10 and 100 km spans. The setup is illustrated in Fig. 6. The wavelength scan span and the sweeping speed of the tunable laser are set to 0.2 nm and 2000 nm/s, respectively. In the probe arm, the light beam is amplified and launched as OAM beam in the same manner as in the VNA setup, where one OAM mode of MG $|l|$ is launched into the RCF at a time. To mitigate the effect of the phase noise of the tunable laser, a 10-km or 100-km SMF reference arm is placed in the interferometer to provide optical path length similar to the probe arm. At the receiver side, all the OAM modes within each MG are converted into Gaussian beams by two commercial vortex phase plates (VPPs) with opposite topological charge signs (+/- $l$). It is noted that, in the probe arm, only one azimuthal OAM mode (+/- $l$) with its both polarizations is detected by an ICR at a time, which is realized by a switch as shown in Fig. 6. Finally, the detected electrical signals are digitized and stored by a 4-channel real-time oscilloscope (OSC) for subsequent off-line processing to obtain the impulse responses. The measurement is repeated for each MG.

The SWI-measured impulse response matrix for all supported OAM modes at 1550 nm of the 10-km RCF transmission system is shown in Fig. 7. Due to the strong coupling among the intra-MG modes, the received impulse responses of all intra-MG modes are required to be detected and used to calculate the inter-MG power coupling coefficients. Compared with the measured impulse response using VNA setup, similar strong distributed mode coupling between OAM MGs $|l| = 0$ and $|l| = 1$ is observed over 10-km RCF system. However, the previously conspicuous plateaus resulting from distributed weak coupling between MGs can now be seen, which is different from that measured using VNA setup.

The main source of error in SWI measurement results from the residual phase noise at the receiver, introduced by the relative delay between the RCF modes in the probe arm and the SMF mode in reference arm [17]. This residual phase noise may lead to varying amount of pulse broadening in the measured impulse responses and make the calculated power coupling coefficients using theoretical-curve fitting approach unreliable, which can be verified from the measured results and analyses in the next section.

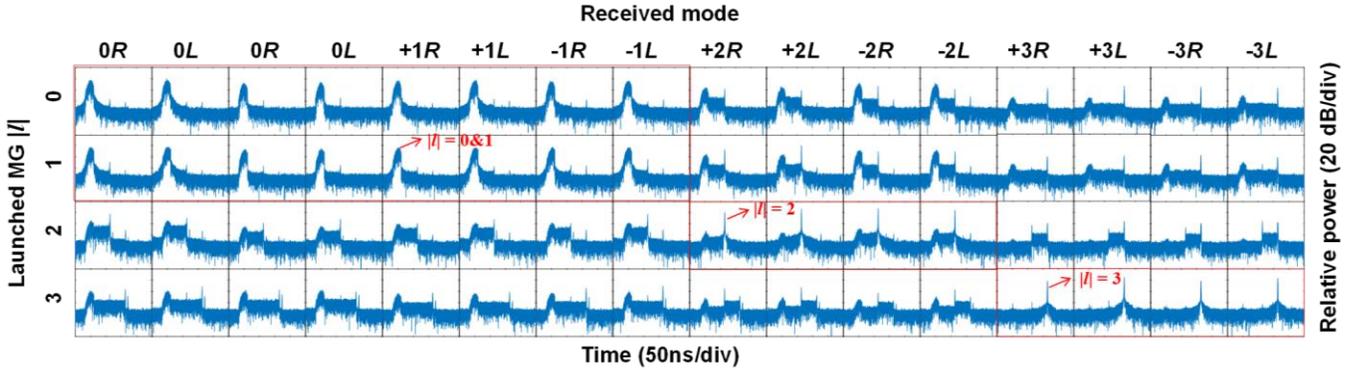

Fig. 7 The SWI-measured impulse response matrix of a 10-km RCF at 1550 nm (where *L* and *R* representing left- and right-handed circular polarizations). Here note that the approximately uniform power distribution between the two discrete modal peaks is resulted from the inter-mode weak coupling.

TABLE II. MEASURED POWER COUPLING COEFFICIENTS BETWEEN ADJACENT HIGH-ORDER MGS AT 1550 NM.

|  | Fiber length (km) | Measured setup | Calculation method | MG | | |
| --- | --- | --- | --- | --- | --- | --- |
|  |  |  |  | $|l| = 1$ | $|l| = 2$ | $|l| = 3$ |
| $h$ (dB/km) | 10 | SWI | Integral calculation | -31.88 |  | -34.36 |
|  | 10 | SWI | Theoretical-curve fitting | -45.7 |  | -46.9 |
|  | 10 | VNA | Integral calculation | -40.5 |  | -41.68 |
|  | 10 | VNA | Theoretical-curve fitting | -52.83 |  | -53.87 |
|  | 100 | SWI | Integral calculation | -33.36 |  | -32.89 |
|  | 100 | SWI | Theoretical-curve fitting | -43.15 |  | -39.86 |
|  | 100 | Power meter | Power ratio | -31.94 |  | -33.27 |

TABLE III. INTEGRAL VALUES OF DISCRETE MODAL PEAKS OR DMGD PLATEAUS OF HIGH-ORDER MODES AT 1550 NM AFTER 10-KM TRANSMISSION.

| Relatively power (dB) | | Received mode | | | | | | | | | | | |
|---|---|---|---|---|---|---|---|---|---|---|---|---|---|
| | | +1R | +1L | -1R | -1L | +2R | +2L | -2R | -2L | +3R | +3L | -3R | -3L |
| Launched MG | $|l|=1$ | 85.79 | 86.72 | 86.08 | 86.68 | 63.17 | 62.08 | 61.54 | 62.24 | 55.57 | 56.26 | 55.15 | 56.38 |
| | $|l|=2$ | 68.47 | 69.49 | 69.06 | 70.23 | 78.35 | 86.51 | 84.98 | 83.35 | 58.8 | 59.78 | 58.48 | 59.66 |
| | $|l|=3$ | 59.39 | 60.39 | 61.13 | 62.3 | 59.55 | 59.78 | 56.83 | 58.16 | 80.19 | 83.76 | 79.67 | 83.9 |

TABLE IV. CALCULATED POWER CROSSTALK BETWEEN HIGH-ORDER MGS USING INTEGRAL VALUES AT 1550 NM AFTER 10-KM TRANSMISSION.

| Relatively power (dB) | | Received MG | | |
|---|---|---|---|---|
| | | $|l|=1$ | $|l|=2$ | $|l|=3$ |
| Launched MG | $|l|=1$ | 0 | -21.88 | -26.44 |
| | $|l|=2$ | -16.98 | 0 | -23.09 |
| | $|l|=3$ | -25.4 | -25.45 | 0 |

## C. Comparisons of evaluated power coupling coefficients using different methods

The power coupling coefficient values evaluated using the integral calculation and theoretical-curve fitting approaches are listed in Table II based on the SWI-measured impulse responses. The integral values of discrete modal peaks or DMGD plateaus of intra-MG modes at 1550 nm after 10-km transmission shown in Table III are summed to calculate the power crosstalk between high-order MGs as the results presented in Table IV. We then obtain the power coupling coefficients between adjacent MGs using Eq. (1). It should be noted that due to the strong coupling between MGs $|l|=0$ and $|l|=1$, the received power of modal peaks for MG $|l|=1$ is reduced by around half (i.e., around 3-dB power loss). Thus we only use the impulse responses (MG $|l|=1$ excitation & MG $|l|=2$ detection) to calculate the power coupling coefficient between MGs $|l|=1$ and $|l|=2$ thanks to the low MDL of the RCF. To avoid the effect of MDL on the measurement accuracy, we sum the integral values of all excited mode peaks and the integral values of all DMGD plateaus belonging to MGs $|l|=2$ and $|l|=3$ before calculating their power coupling coefficient. The evaluated power coupling coefficients based on the VNA-measured impulse responses are also presented in Table II for comparison.

To verify the reliability of these methods, we also measured the power coupling coefficients in the 100-km RCF by both the abovementioned methods and directly measuring the total power crosstalk of the whole system considering both the power coupling from the RCF and the Mux/ DeMux, as the result shown in Table II.

From Table II, the following observations can be made of the power coupling coefficients:

1) The values obtained by the theoretical-curve fitting approach are much lower than that by the integral calculation approach due to abovementioned sources of errors in section II.

2) The values obtained by the VNA setup are lower than those by the SWI setup and the direct power measurement, due to the suppression of the distributed weak coupling in the VNA setup.

3) For both 10-km and 100-km RCF spans, the values obtained by the SWI setup and integral-calculation approach exhibit similar values, respectively ranging from -31.88 to -33.36 dB/km for MGs $|l|=1$ & 2 and -32.89 to -34.36 dB/km for MGs $|l|=2$ & 3, which are close to the values of -31.94 dB/km and -33.27 dB/km measured by power meter in the 100-km RCF system. Direct crosstalk power measurement over very long fiber span is justifiable because the accumulated distributed mode coupling is dominant compared to the discrete mode coupling at the Mux/ DeMux. This has been verified by the directly measured power coupling in a 1-km RCF transmission system, where crosstalk of around -24 dB between the high-order MGs $|l|=2$ and 3 have been obtained which is dominated by crosstalk induced by the Mux/ DeMux [12]. This is much lower than the values of -13.27 dB after 100 km transmission. The -13.27 dB value in turn is marginally (~1 dB) lower than that reported in [14] thanks to the further optimization of the Mux/ DeMux.

Therefore, it can be concluded that the combination of integral calculation approach and SWI measurement produces the most reliable coupling coefficient values.

## D. C-Band Power Coupling Coefficient Evaluation Using Integral Calculation Approach based SWI Measurement

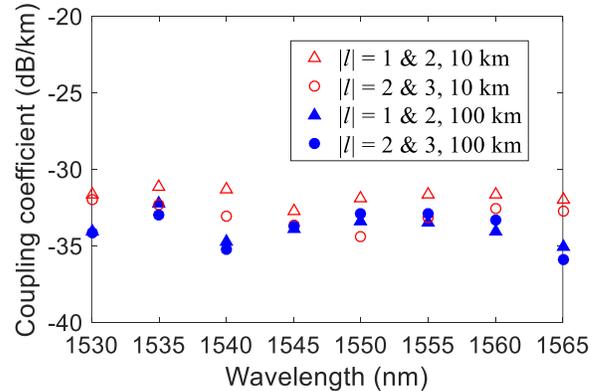

Fig. 8 The measured power coupling coefficient versus wavelength for SWI measurement based 10-km and 100-km RCF systems.

Using the SWI measurement combined with the integral calculation approach, the power coupling coefficients over the entire C band for both the 10-km and 100-km RCF spans are evaluated, as the results shown in Fig. 8. Low power coupling coefficients of < -31 dB/km and <-32 dB/km between adjacent high-order OAM MGs after 10-km and 100-km fiber propagation within the entire C band have been confirmed, respectively. The slightly lower values of power coupling coefficients in the 100-km fiber span compared to the 10-km fiber span could be attributed to the relatively lower inter-modal Rayleigh forward-scattering at longer propagation distance [18].

## IV. CONCLUSIONS

In this paper, we have proposed a modified inter-mode coupling characterization method based on SWI impulse-response measurement and integral calculation. The proposed method eliminates significant sources of error that may lead to underestimation of the power coupling coefficients and has been verified by direct crosstalk power measurement over 100-km long fiber span that allows the Mux/ Demux crosstalk to be discounted. Based on the proposed characterization method, low average power coupling coefficients of <-31 dB/km and <-32 dB/km between adjacent high-order OAM MGs (topological charge $|l|$ = 1 & 2, 2 & 3) over 10-km and 100-km RCF spans within the entire C band have been achieved, respectively.

## APPENDIX

The power coupling coefficient $h$ between a pair of adjacent high-order modes (mode 1 and mode 2) can be evaluated based on measured impulse responses after fiber transmission, when exciting either mode 1 or 2 using the following equations [8], [16].

For mode 1 excitation:

$$P_{11}(z,\tau) = \begin{cases} \delta(\tau)\exp\{-(\alpha_1+h)z\}, \tau = 0 \\ \delta(\tau)\exp\{-(\alpha_1+h)z\} + \\ \left\{\frac{hn_0}{c\tau_g}\sqrt{\frac{1-\tau}{\tau}}I_1(X')\right\}\exp\{-(\alpha_1+h)z\}\exp\{(\alpha_1-\alpha_2)\tau z\}, 0 < \tau \leq 1 \end{cases}$$

(2)

$$P_{12}(z,\tau) = \left\{\frac{hn_0}{c\tau_g}I_0(X')\right\}\exp\{-(\alpha_1+h)z\}\exp\{(\alpha_1-\alpha_2)\tau z\}, 0 < \tau \leq 1$$

(3)

For mode 2 excitation:

$$P_{21}(z,\tau) = \left\{h(1+\frac{n_0}{c\tau_g})I_0(X')\right\}\exp\{-(\alpha_1+h)z\}\exp\{(\alpha_1-\alpha_2)\tau z\}, 0 \leq \tau < 1$$

(4)

$$P_{22}(z,\tau) = \begin{cases} \delta(1-\tau)\exp\{-(\alpha_2+h)z\} + \\ \left\{h(1+\frac{n_0}{c\tau_g})\sqrt{\frac{\tau}{1-\tau}}I_1(X')\right\}\exp\{-(\alpha_1+h)z\}\exp\{(\alpha_1-\alpha_2)\tau z\} \\ , 0 \leq \tau < 1 \\ \delta(1-\tau)\exp\{-(\alpha_2+h)z\}, \tau = 1 \end{cases}$$

(5)

$$\tau \equiv \frac{t-t_1}{\tau_g}$$

(6)

$$X' \equiv 2hz\sqrt{\tau(1-\tau)}$$

(7)

where $P_{i,j}$ is the received relative power distribution in the case of mode $i$ excitation and mode $j$ detection. $\alpha_i$ is the attenuation of mode $i$ ($i$ = 1 or 2), $c$ is velocity of light in vacuum, $n_0$ is the maximum index of the core, $\tau_g$ is DMGD between two modes, $z$ is the fiber length, $\tau$ is the normalized delay time calculated according to Eq. (6), where $t_1$ is the relatively lower time delay of the two modes. $I_p$ ($p$ = 0, 1) is the modified Bessel function of the first kind for order $p$. $\delta(x)$ is a delta function. According to the theoretical impulse responses for two-mode transmission reported in [16], here we correct the error by replacing the term $\exp\{-(\alpha_2+h)z\}$ in Eqs. (4) and (5) presented in [8] with $\exp\{-(\alpha_1+h)z\}$.

## REFERENCES

[1] D. J. Richardson, J. M. Fini, and L. E. Nelson, "Space-division multiplexing in optical fibres," *Nat. Photonics*, vol. 7, no. 5, pp. 354-362, Apr. 2013.

[2] G. Li, N. Bai, N. Zhao, and C. Xia, "Space-division multiplexing: the next frontier in optical communication," *Adv. Opt. Photon.*, vol. 6, no. 4, pp. 413-487, Dec. 2014.

[3] G. Zhu, Z. Hu, X. Wu, C. Du, W. Luo, Y. Chen, X. Cai, J. Liu, J. Zhu, and S. Yu, "Scalable mode division multiplexed transmission over a 10-km ring-core fiber using high-order orbital angular momentum modes," *Opt. Express*, vol. 26, no. 2, pp. 594-604, Jan. 2018.

[4] S. Arık, D. Askarov, and J. M. Kahn, "Effect of Mode Coupling on Signal Processing Complexity in Mode-Division Multiplexing," *J. Lightwave Technol.*, vol. 31, no. 3, pp. 423-431, Feb. 2013.

[5] N. K. Fontaine, R. Ryf, M. A. Mestre, B. Guan, X. Palou, S. Randel, Y. Sun, L. Grüner-Nielsen, R. V. Jensen, and R. Lingle, "Characterization of Space-Division Multiplexing Systems using a Swept-Wavelength Interferometer," in *Proc. Opt. Fiber Commun.*, 2013, Paper OW1K.2.

[6] N. K. Fontaine, "Characterization of Space-Division Multiplexing Fibers using Swept-Wavelength Interferometry," in *Proc. Opt. Fiber Commun.*, 2015, Paper W4I.7.

[7] S. Jiang, L. Ma, Z. Zhang, X. Xu, S. Wang, J. Du, C. Yang, W. Tong, and Z. He, "Design and Characterization of Ring-Assisted Few-Mode Fibers for Weakly Coupled Mode-Division Multiplexing Transmission," *J. Lightwave Technol.*, vol. 36, no. 23, pp. 5547-5555, Oct. 2018.

[8] R. Maruyama, N. Kuwaki, S. Matsuo, and M. Ohashi, "Relationship Between Mode Coupling and Fiber Characteristics in Few-Mode Fibers Analyzed Using Impulse Response Measurements Technique," *J. Lightwave Technol.*, vol. 35, no. 4, pp. 650-657, Sep. 2017.

[9] C. Shi, L. Shen, J. Zhang, J. Liu, L. Zhang, J. Luo, J. Liu, and S. Yu, "Ultra-Low Inter-Mode-Group Crosstalk Ring-Core Fiber Optimized Using Neural Networks and Genetic Algorithm," in *Proc. Opt. Fiber Commun.*, 2020, Paper W1B.3.

[10] F. Feng, X. Jin, D. O'Brien, F. Payne, Y. Jung, Q. Kang, P. Barua, J. K. Sahu, S.-u. Alam, D. J. Richardson, and T. D. Wilkinson, "All-optical mode-group multiplexed transmission over a graded-index ring-core fiber with single radial mode," *Opt. Express*, vol. 25, no. 12, pp. 13773-13781, Jun. 2017.

[11] J. Zhang, G. Zhu, J. Liu, X. Wu, J. Zhu, C. Du, W. Luo, Y. Chen, and S. Yu, "Orbital-angular-momentum mode-group multiplexed transmission over a graded-index ring-core fiber based on receive diversity and maximal ratio combining," *Opt. Express*, vol. 26, no. 4, pp. 4243–4257, Feb. 2018..

[12] J. Zhang, J. Liu, Z. Lin, J. Liu, L. Shen, and S. Yu, "Nonlinearity-Aware Adaptive Bit and Power Loading DMT Transmission Over Low-Crosstalk Ring-Core Fiber With Mode Group Multiplexing," *J. Lightwave Technol.*, vol. 38, no. 21, pp. 5875-5882, Jun. 2020.

[13] H. Wang, Y. Liang, X. Zhang, S. Chen, L. Shen, L. Zhang, J. Luo, and J. Wang, "Low-Loss Orbital Angular Momentum Ring-Core Fiber: Design, Fabrication and Characterization," *J. Lightwave Technol.*, vol. 38, no. 22, pp. 6327-6333, Jun. 2020.

[14] J. Zhang, J. Liu, L. Shen, L. Zhang, J. Luo, J. Liu, and S. Yu, "Mode-division multiplexed transmission of wavelength-division multiplexing signals over a 100-km single-span orbital angular momentum fiber," *Photon. Res.*, vol. 8, no. 7, pp. 1236-1242, Jul. 2020.

[15] L. Shen, J. Zhang, J. Liu, G. Zhu, Z. Lin, Y. Luo, Z. Luo, L. Zhang, J. Luo, C. Guo, J. Liu, and S. Yu, "MIMO-free WDM-MDM transmission over 100-km single-span ring-core fibre," in *Proc. Eur. Conf. Opt. Commun.*, 2019, pp. 1-4.
[16] S. Kawakami and M. Ikeda, "Transmission characteristics of a two-mode optical waveguide," *IEEE J. Quantum Electron.*, vol. 14, no. 8, pp. 608-614, 1978.
[17] X. Fan, Y. Koshikiya, and F. Ito, "Phase-Noise-Compensated Optical Frequency-Domain Reflectometry," *IEEE J. Quantum Electron.*, vol. 45, no. 6, pp. 594-602, 2009.
[18] Z. Wang, H. Wu, X. Hu, N. Zhao, Q. Mo, and G. Li, "Rayleigh scattering in few-mode optical fibers," *Scientific Reports*, vol. 6, no. 1, pp. 35844, Oct. 2016.